\documentclass[a4paper,11pt]{article}
\usepackage{comment}
\usepackage{graphicx} 
\usepackage[utf8]{inputenc}
\usepackage[margin=1.0in]{geometry}
\usepackage{authblk}
\usepackage{setspace}
\usepackage{lineno}
\usepackage{caption}
\usepackage{subcaption}
\usepackage{amsmath}
\usepackage{lipsum}
\usepackage{wrapfig}

\title{\bf Sensitivity Study of Supernova Neutrinos for Mass Hierarchy}

\author[a*]{Riya Gaba}
\author[a]{Maitreyee Mukherjee}
\author[a]{Vipin Bhatnagar}

\affil[a]{Department of Physics, Panjab University, Chandigarh, India}

\affil[*]{Address correspondence to: riyagaba@fnal.gov}

\begin{document}
\maketitle

\begin{abstract}
    
Supernovae represent some of the most energetically explosive events in the universe, with a substantial fraction of their released gravitational energy carried away by neutrinos. This study evaluates the sensitivity of three next-generation neutrino detectors which are Deep Underground Neutrino Experiment (DUNE), Hyper-Kamiokande (Hyper-K), and the Jiangmen Underground Neutrino Observatory (JUNO) to supernova neutrinos, specifically focusing on their ability to discern between normal and inverted mass hierarchies. We utilize three different flux models: Bollig, Tamborra and Nakazato, employing the SNEWPY software to simulate the expected neutrino fluxes from core-collapse supernovae. These models highlight the variability in the predictions due to different progenitors and simulation methods. By analyzing the event rates across various interaction channels and implementing the adiabatic Mikheyev-Smirnov-Wolfenstein (MSW) effect on the neutrino flux, we have calculated the expected detection rates for each detector. Our results indicate that sensitivities range from ${\sim}3\sigma$ to ${\sim}9\sigma$ for DUNE, ${\sim}4\sigma$ to ${\sim}16\sigma$ for Hyper-K and ${\sim}1.7\sigma$ to ${\sim}6.7\sigma$ for JUNO depending on distance and flux model, with sensitivity diminishing significantly at larger distances. This work underscores the potential of future neutrino observatories to enhance our understanding of fundamental physics through the study of supernova neutrinos.

\end{abstract}

\flushbottom

\section{Introduction}

When a massive star reaches the end of its life cycle, it can undergo a dramatic and violent explosion known as a supernova. This cosmic event not only marks the death of a star but also provides a unique opportunity for scientists to study the fundamental processes of the universe. One of the most intriguing aspects of a supernova is the emission of neutrinos—nearly massless, electrically neutral particles that are notoriously difficult to detect yet carry crucial information about the stellar explosion.

The few dozen recorded neutrino events from SN1987A \cite{a} which was observed in Large Magellanic Cloud, about 50 kpc from earth, which is the only supernova observed so far, heralded the era of neutrino astronomy\cite{b}. The events recorded have confirmed the basic physical picture of core collapse and yielded constraints on a wide range of new physics. The community anticipates much more data and corresponding advances in knowledge when the next nearby star collapses.

As our understanding of supernovae advances, next-generation neutrino detectors are poised to revolutionize our ability to observe and analyze these fleeting particles. Among the leading facilities designed to detect neutrinos in the near future are the Deep Underground Neutrino Experiment (DUNE)\cite{c}, Hyper-Kamiokande (Hyper-K)\cite{d}, and the Jiangmen Underground Neutrino Observatory (JUNO)\cite{e}.

Together, these advanced detectors represent a significant leap forward in neutrino astronomy. By capturing and analyzing supernova neutrinos, they will not only enhance our understanding of stellar explosions but also test fundamental physics theories, refine models of stellar evolution, and potentially uncover new phenomena in the realm of particle physics. The next decade promises exciting advancements as these cutting-edge facilities begin their operations and contribute to one of the most profound areas of astrophysics and cosmology\cite{f}.

\section{Supernova neutrinos}

Neutrinos carry about 99\% of the gravitational energy released during supernova explosion which is estimated to be about $10^{53}$ ergs\cite{a}.
The average energy of the emitted neutrinos is around 10 MeV and approximately $10^{58}$ neutrinos are emitted during the core-collapse supernova. Core-collapse SNe is the final explosion of single stars with a mass at least 8-10 solar mass($M_\odot$).
Despite their abundance, neutrinos interact very weakly with matter, making them extremely challenging to observe. Specialized detectors, often located deep underground to shield them from other cosmic rays and particles, are employed to capture and analyze these fleeting signals.

\begin{itemize}
    \item {\bf Core-Collapse Dynamics:}
            Stars with the mass of about 10 $M_\odot$ are considered to undergo all stages of nuclear fusion until the star forms an onion like structure with an iron core surrounded by shells of decreasing atomic masses. The gravitational force is balanced by degenerate electron pressure but iron being a tightly bound nucleus cannot further be burned \cite{g}. The core starts contracting and some stages later it undergoes explosion. Neutrinos of all flavors are produced\cite{h} in this process through processes such as pair annihilation, electron-nucleon bremsstrahlung, nucleon-nucleon bremsstrahlung, plasmon decay, and photoannhilation.
    \item {\bf Matter effects:}
        As we know neutrinos have mass and their three flavors mix, the neutrino fluxes from supernovae will be influenced by oscillation parameters. Additionally, matter oscillation effects (known as Mikheyev-Smirnov-Wolfenstein effects) will affect neutrino fluxes, where neutrinos experience matter potential\cite{i} given by $\lambda = \sqrt{2}G_fn_e(r)$, $G_f$ and $n_e$ being fermi constant and electron density respectively. This will alter the spectra as neutrinos pass through dense matter. Hence, the characteristics of neutrino spectra and their temporal changes are influenced by mass and oscillation parameters, such as $\theta_{13}$ and the mass hierarchy. Specifically, the spectra of $\nu_e$ and $\bar{\nu_e}$ will vary which could help determine the hierarchy\cite{j,k,l}.
        This propagation of neutrinos in matter can be adiabatic or non-adiabatic.
         
\end{itemize}

\section{Flux Models}

Despite significant advances in computer simulations of core-collapse supernovae over recent decades, these simulations remain constrained by current computing power. To address this limitation, modeling groups use various approximations and simplifying assumptions.

To demonstrate the wide applicability of the models, we use a varied selection of three distinct models. This includes: a one-dimensional model of mostly historical significance, a one-dimensional model from recent parametric studies, and a complex multi-dimensional model. These simulations, carried out by different teams using various progenitors and simulation codes, aim to represent the broad range of available models.

To study different flux models, the open-sourced software package SNEWPY(Supernova Neutrino Early Warning Models for Python)\cite{m} is used. The models under consideration are, following SNEWPY’s nomenclature, Bollig 2016(11.2$M_\odot$) \cite{g}, Tamborra 2014 (11.2$M_\odot$)\cite{n}, Nakazato 2013(13$M_\odot$ )\cite{o}. This set of models aims to reflect the variability in the existing neutrino predictions among different models, computational approaches and progenitor masses. To account for flavor transformations, the neutrino flux predictions are modified according to AdiabaticMSW transformations using SNEWPY. These transformations depend on $\theta_{13}$ and $\theta_{23}$ \cite{p}, with values chosen from the Particle Data Group\cite{q}. Neutrino flux models have the neutrino luminosity divided into four flavor categories: $L_{\nu_e}$ , $L_{\Bar{\nu_e}}$ , $L_{\nu_x}$ and $L_{\Bar{\nu_e}}$ , where $x = \mu + \tau$. 

\begin{figure}
    \centering
    \hspace*{-4.0em}
    \includegraphics[width=0.32\linewidth]
    {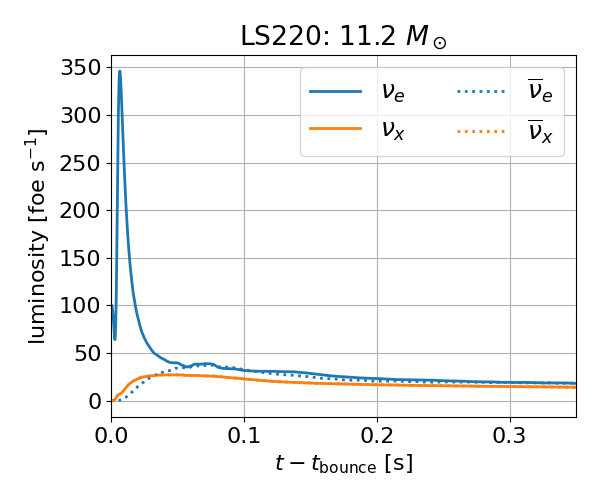}
    \qquad
    \hspace*{-2.0em}
    \includegraphics[width=0.32\linewidth]{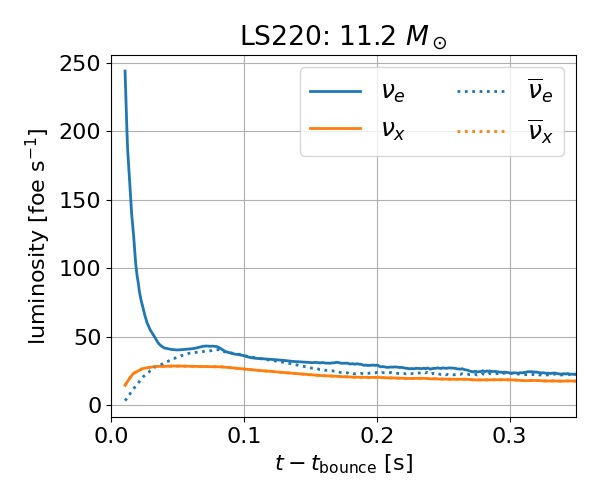}
    \qquad
    \hspace*{-2.0em}
    \includegraphics[width=0.32\linewidth]{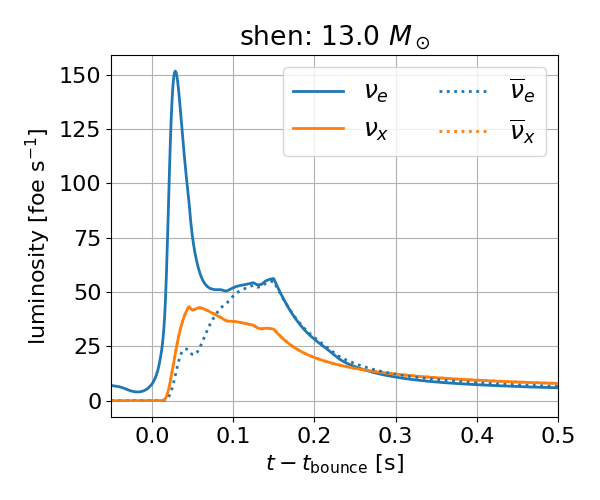}
    \caption{Luminosity predicted by simulation for progenitor of solar mass 11.2$M_\odot$ in case of Bollig and Tamborra and for 13$M_\odot$ in case of Nakazato}
    \label{models_lum}
\end{figure}

\begin{figure}
    \centering
    \includegraphics[width=1.0\linewidth]{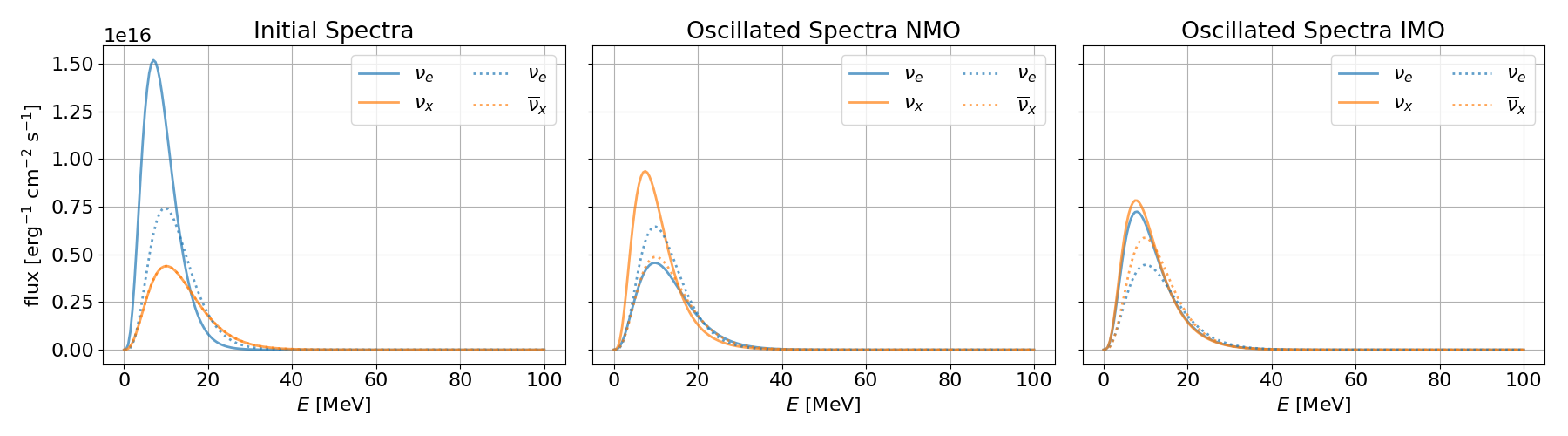}
    \caption{Flux according to Bollig supernova model integrated for 10s}
    \label{fig:Bollig_flux}
\end{figure}

\begin{figure}
    \centering
    \includegraphics[width=1.0\linewidth]{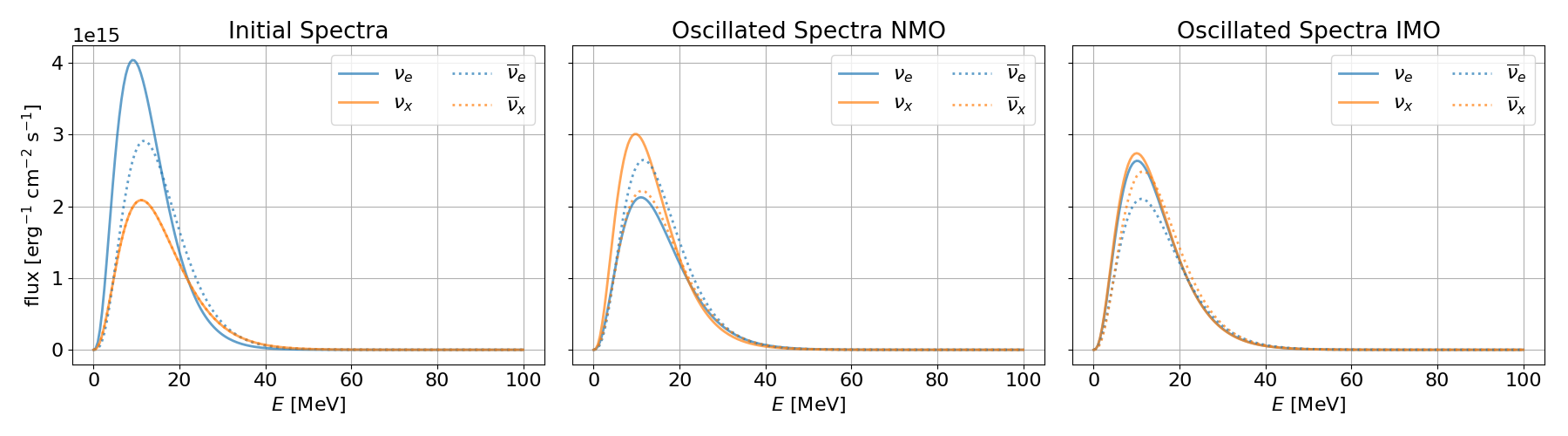}
    \caption{Flux according to Tamborra supernova model integrated for 10s}
    \label{fig:Tamborra_flux}
\end{figure}

\begin{figure}
    \centering
    \includegraphics[width=1.0\linewidth]{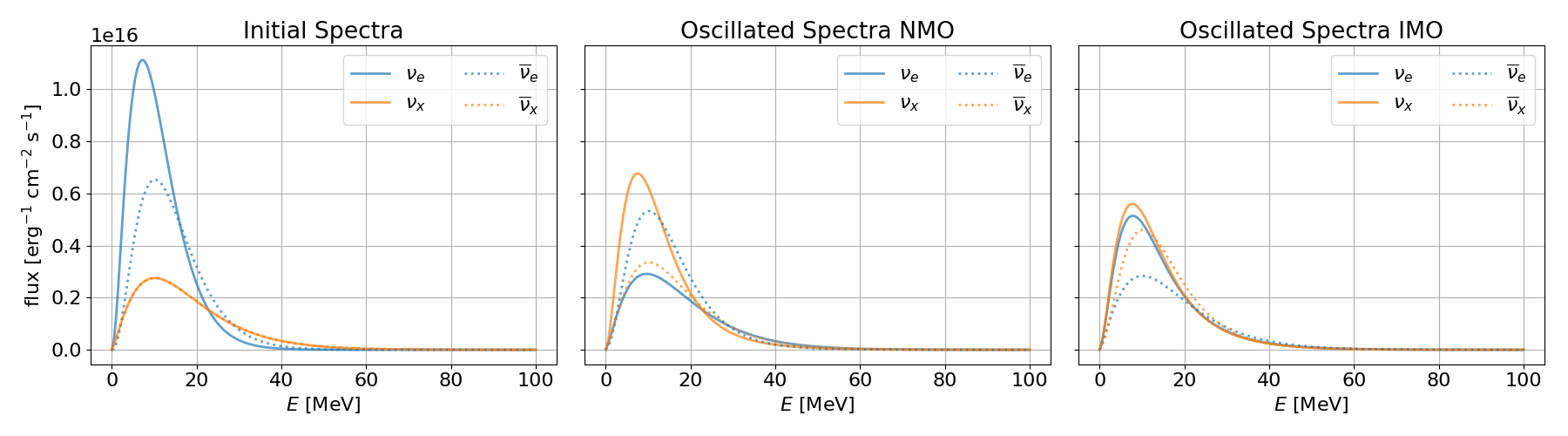}
    \caption{Flux according to Nakazato supernova model integrated for 10s}
    \label{fig:Nakazato_flux}
\end{figure}

Among models used here, Bollig is spherically symmetric 1D model. It utilizes a two-moment scheme for the transport of neutrinos and antineutrinos of all three flavors and accounts for the full energy and velocity (to order (v/c)) dependence of the transport in the co-moving frame of the fluid. Luminosity for this model is shown in fig.\ref{models_lum} and flux used to calculate event rate is shown in fig.\ref{fig:Bollig_flux}.
However, Tamborra model \cite{r}\cite{n} is a pioneering three-dimensional supernova simulation with energy dependent neutrino transport. We use results from the simulation of a $11.2 M_\odot$ progenitor\cite{s}. The simulation was performed using the Prometheus-Vertex code, and flux used to calculate event rate is shown in fig.\ref{fig:Tamborra_flux} where flux before and after the msw transformation is shown.
For Nakazato\cite{o}, family of models (Nakazato et al. 2013) contains progenitors with different initial masses and metallicities. In this work, we focus on the 13 $M_\odot$ progenitor with solar metallicity (Z = 0.02). Here, we use data from the first stage of the simulation, which contains the first 520 ms post bounce. Luminosity for this model is shown in fig.\ref{models_lum} and flux used to calculate event rate for this model is shown in fig.\ref{fig:Nakazato_flux}.

\section{Interaction channels and event rates}

SNEWPY\cite{m} package provides access to an extensive library of supernova models and flavor transformations, connecting the neutrino fluxes generated in supernovae to those observed on Earth. Additionally, it includes a Python interface to SNOwGLoBES\cite{w}, which computes event rate by folding input fluxes with cross-sections and detector parameters. The output is in the form of “interaction” rates for each neutrino interaction channel as a function of incident neutrino energy, and “smeared” rates as a function of detected energy for each channel (i.e. the spectrum that actually will be observed in a detector). SNOwGLoBES employs the smearing matrices to translate incident neutrino energies from the flux file into the energies deposited in the detector as seen in the simulation results. 

Here we carried out study for three future detectors namely, DUNE, Hyper-k, JUNO which although different in their target material, are useful to cover wide portion of flux giving us a bigger picture. 

In our study, the energy range of supernova neutrinos is taken from 5 to 100 MeV, with 200 energy bins for both true and test event rate calculation. Background is not considered here as it is negligible due to detectors being underground. For systematic uncertainties, we have considered $5\%$ error in normalization and energy calibration in detectors. We assume oscillation parameters, energy resolution and efficiencies as provided in \cite{w}.

\begin{table}[]
    \centering
    \begin{tabular}{|c|c|c|c|}
    \hline
  \bf SN Model & \bf DUNE & \bf Hyper-k & \bf JUNO \\
      \hline    
   \bf Bollig & 2696/2509 & 27570/27735 & 3767/3783 \\
     \hline
    \bf Tamborra & 761/732 & 7998/7564 & 1080/1023 \\
     \hline
    \bf Nakazato & 3405/3059 & 24203/29922 & 3433/4112 \\
     \hline
    \end{tabular}
    \caption{Total event rate for above mentioned SN flux model after adiabatic MSW flavor transformation is applied where numbers on the left indicates normal mass ordering and inverse mass ordering on the right}
    \label{Table:event rate}
\end{table}

\begin{itemize}
    \item {\bf DUNE:}
    Located deep underground in the Sanford Underground Research Facility in South Dakota, DUNE is set to be one of the most advanced neutrino experiments ever built. It features a massive liquid argon time projection chamber (LArTPC), which offers exceptional sensitivity and resolution for detecting neutrinos. In the context of supernovae\cite{x}, as shown in Table \ref{Table:event rate}, we can see that DUNE expects around total of 3000 events for two of flux models and around 1000 for one , and DUNE's capabilities will allow for precise measurement of the neutrino burst's energy and flavor composition, offering insights into the explosion dynamics and core processes of supernovae. Liquid argon has a particular sensitivity to the $\nu_e$ component of a supernova neutrino burst, via the dominant interaction, CC absorption of $\nu_e$ on ${}^{40}Ar$\cite{y}\cite{z}. Additional channels include a $\Bar{\nu_e}$ CC interaction and ES on electrons\cite{aa}. Cross sections for the most relevant interactions are shown in fig.\ref{xscn_argon}. The event rate distribution for DUNE are shown in fig.\ref{fig:eventRateDune}.

\begin{figure}
    \centering
    \includegraphics[width=0.4\linewidth,trim=0 0.5cm 0.1cm 0,clip]{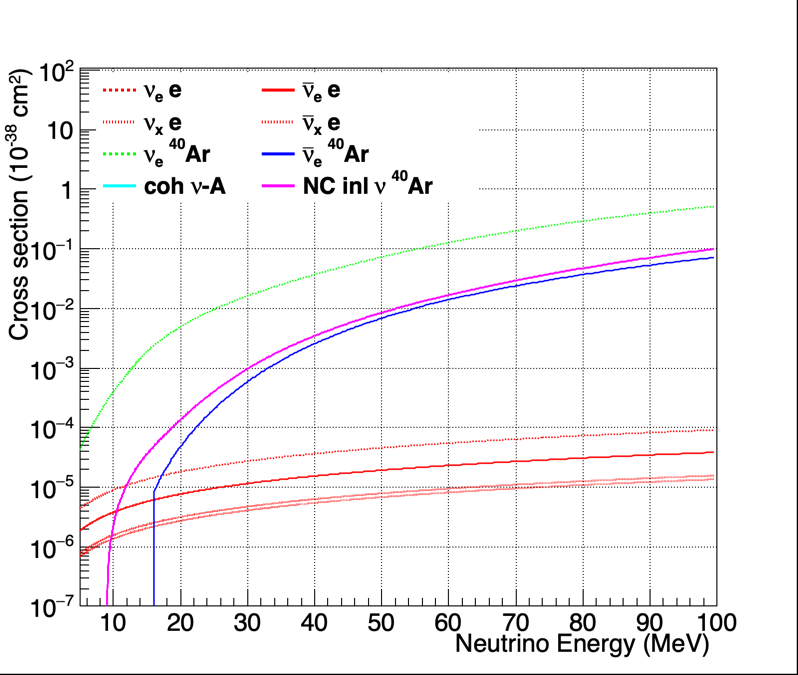}
    \caption{Cross-sections for relevant interaction channels of argon are shown.}
    \label{xscn_argon}
\end{figure}

\begin{figure}
    \centering
    \includegraphics[width=7.5cm]{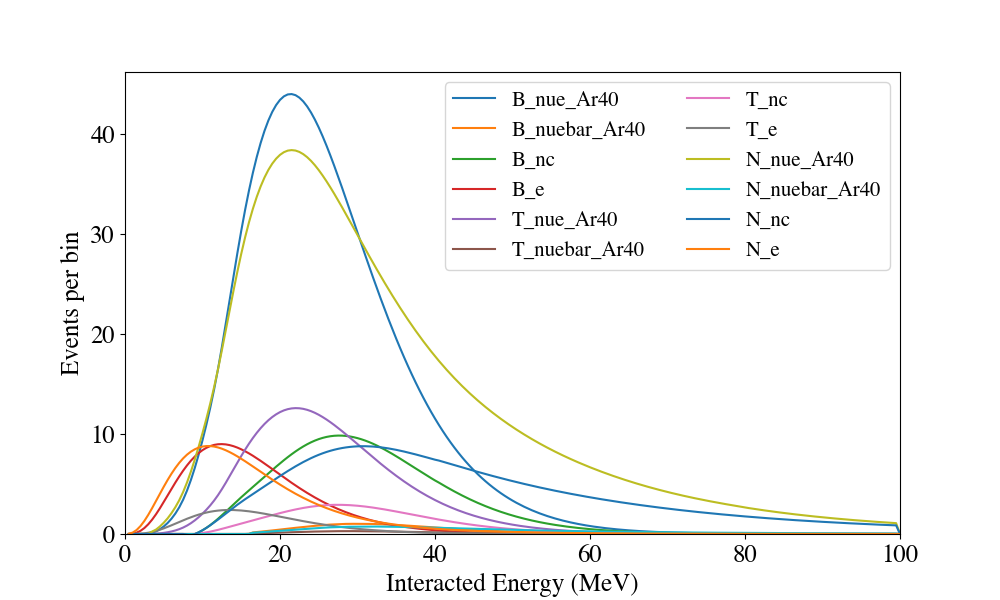}
    \hspace*{-3.8em}
    \qquad
   \includegraphics[width=7.5cm]{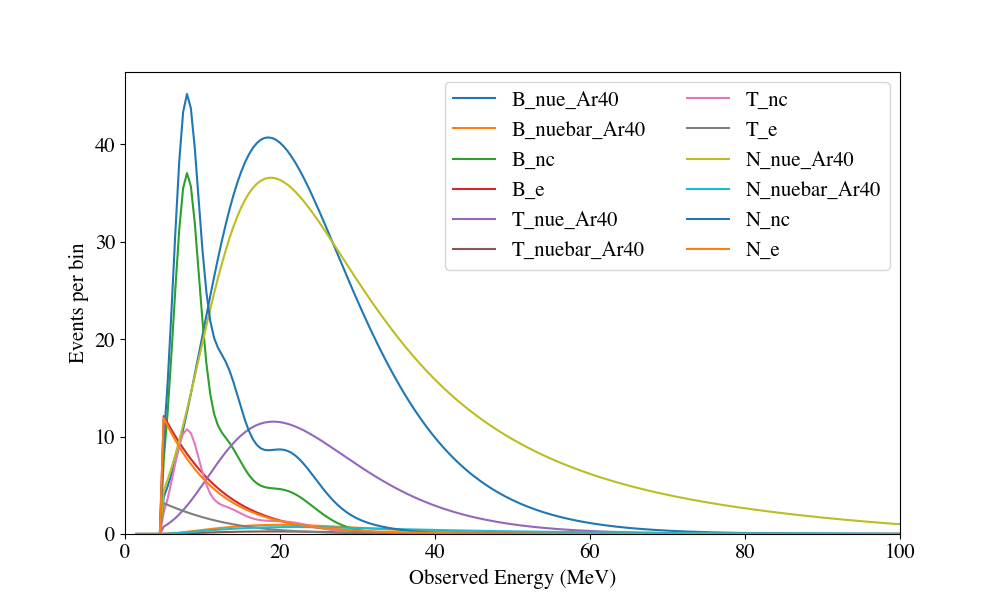}
    \caption{Event rates of DUNE where figure on the left shows us the event rates without smearing being applied whereas the right one shows event rates with smearing}
    \label{fig:eventRateDune}
\end{figure}

    \item {\bf Hyper-Kamiokande}
        Hyper-K is a proposed next-generation water cherenkov detector \cite{bb} located in Japan. Hyper-K's large detector volume as compared to Super-k will increase its chances of capturing neutrinos from a supernova in our galaxy, providing critical data on the neutrino spectrum and time evolution of the burst. Hyper-Kamiokande can reconstruct the time and energy of each individual event, allowing it to reconstruct the neutrino spectrum. Hyper-k expects around 27k events total events according to Bollig model, 8k according to Tamborra and around 24k according to Nakazato supernova model as shown in Table \ref{Table:event rate}. The main interaction channel, inverse beta decay is responsible for about 90\% of events, making Hyper-Kamiokande most sensitive to $\Bar{\nu_e}$. Here cross-section model from reference \cite{cc} is used, shown in fig.\ref{xscn_water} . Elastic neutrino-electron scattering\cite{aa} is a subdominant interaction channel to which all neutrino flavours contribute. Charged-current interactions of $\nu_e$ and $\Bar{\nu_e}$ on $^{16}O$ nuclei are subdominant channels which is used from reference\cite{dd}.
\begin{figure}
    \centering
    \includegraphics[width=0.4\linewidth,trim=0 0.5cm 0.1cm 0,clip]{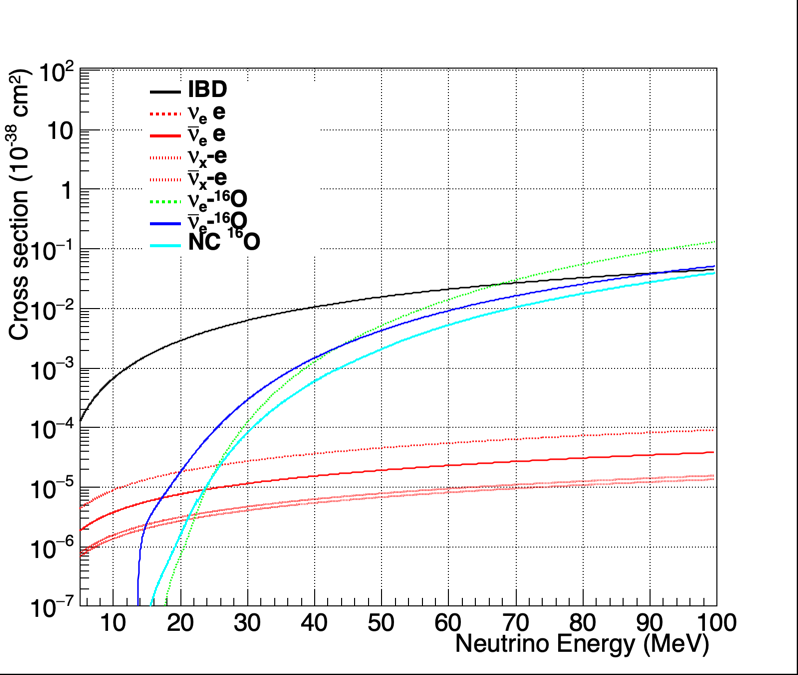}
    \caption{Cross-sections for relevant interaction channels of water are shown.}
    \label{xscn_water}
\end{figure}

\begin{figure}
    \centering
   \includegraphics[width=7.5cm]{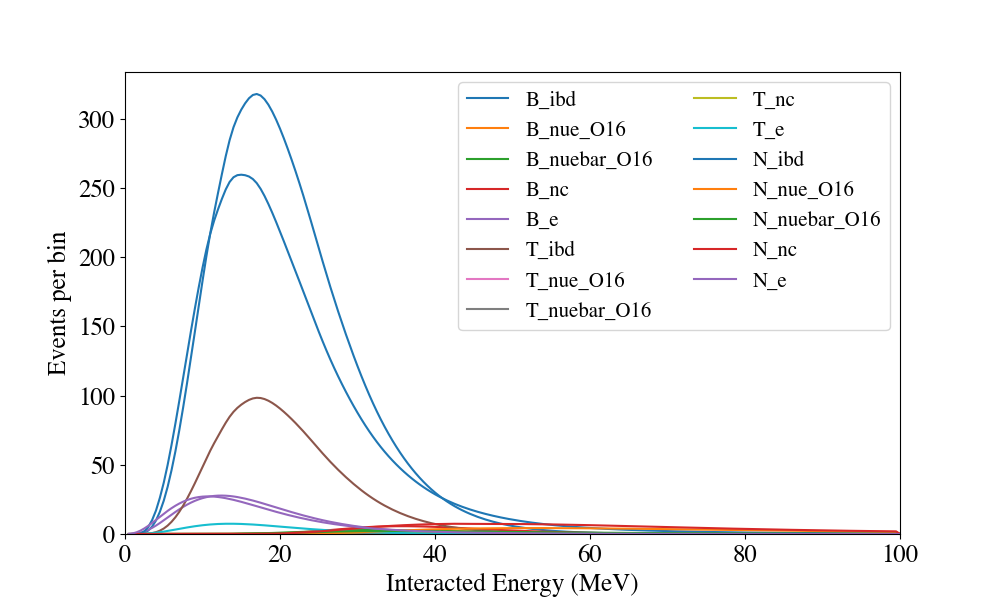}
    \hspace*{-3.8em}
    \qquad
    \includegraphics[width=7.5cm]{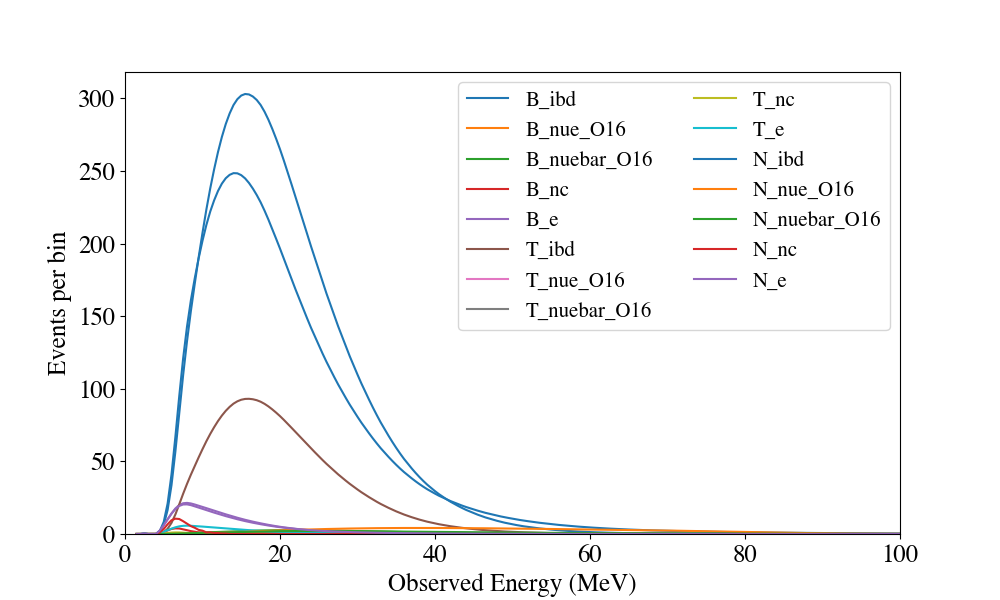}
    \caption{Event rates of Hyper-k where the left plot shows us the event rates without smearing being applied whereas the right plot shows event rates with smearing}
    \label{fig:EventRateHyperk}
\end{figure}

    \item {\bf JUNO:}
        Situated in China, the Jiangmen Underground Neutrino Observatory\cite{e} will feature a large liquid scintillator detector. As the largest new generation LS detector, JUNO will be superior in its high statistics and flavor informations. Since the average SN distance is around 10 kpc, JUNO expect to register about 4k events for two of the flux models and about 1k for the third one like shown in Table \ref{Table:event rate}. The main interaction channel is inverse beta decay(IBD)\cite{cc} caused by the interaction of electron antineutrinos with the LS, other channels include events from all-flavour elastic neutrino-proton elastic scattering,  neutrino-electron elastic scattering as well as other charged current (CC) where cross-section is used from \cite{ee} and neutral current (NC) interactions on the $^{12}C$ nuclei whose cross-section is taken from reference\cite{ff}.

\begin{figure}
    \centering
    \includegraphics[width=0.4\linewidth,trim=0 0.5cm 0.1cm 0,clip]{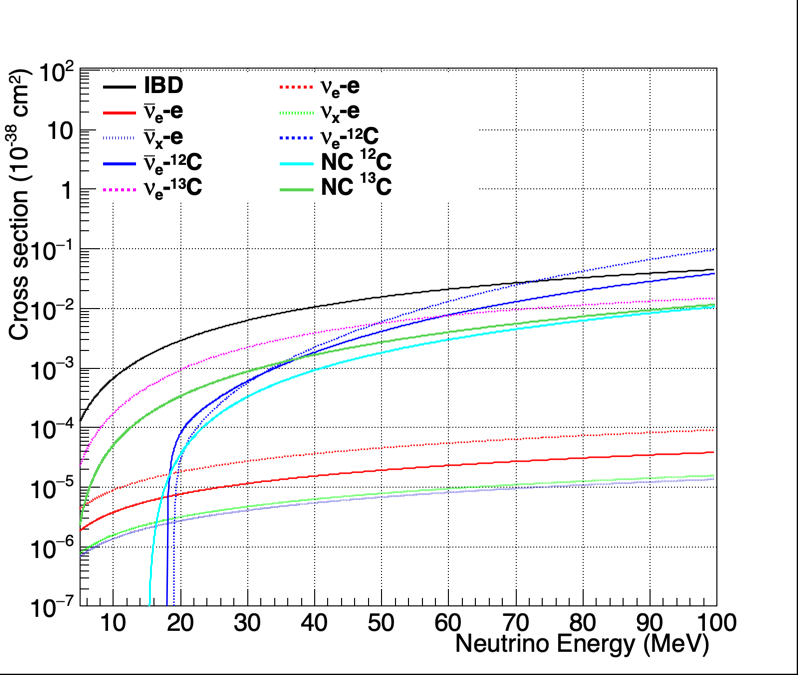}
    \caption{Cross-sections for relevant interaction channels of liquid scintillator are shown.}
    \label{xscn_scint}
\end{figure}

\begin{figure}
    \centering
    \includegraphics[width=7.5cm]{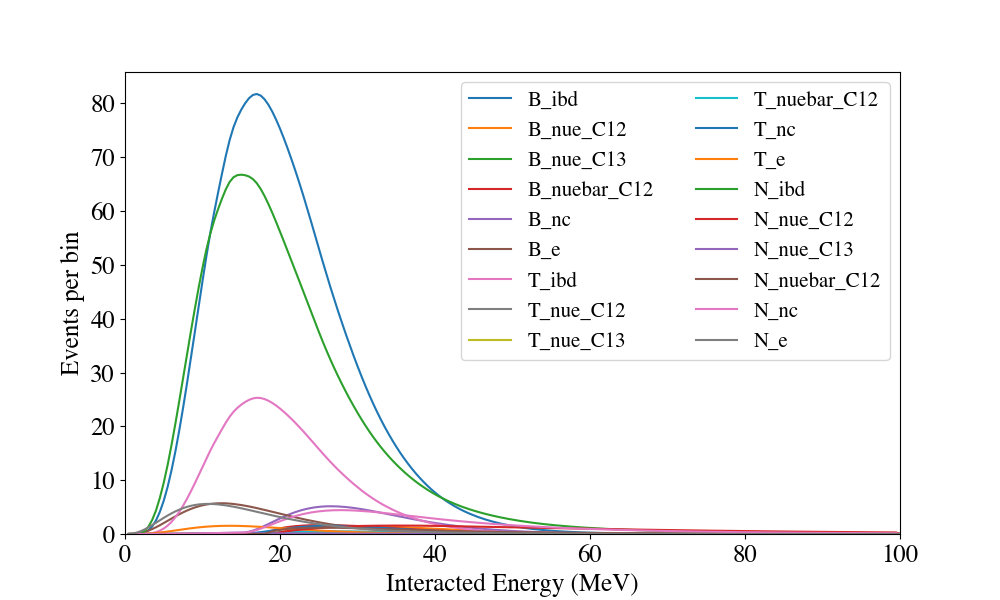}
    \hspace*{-3.8em}
    \qquad
    \includegraphics[width=7.5cm]{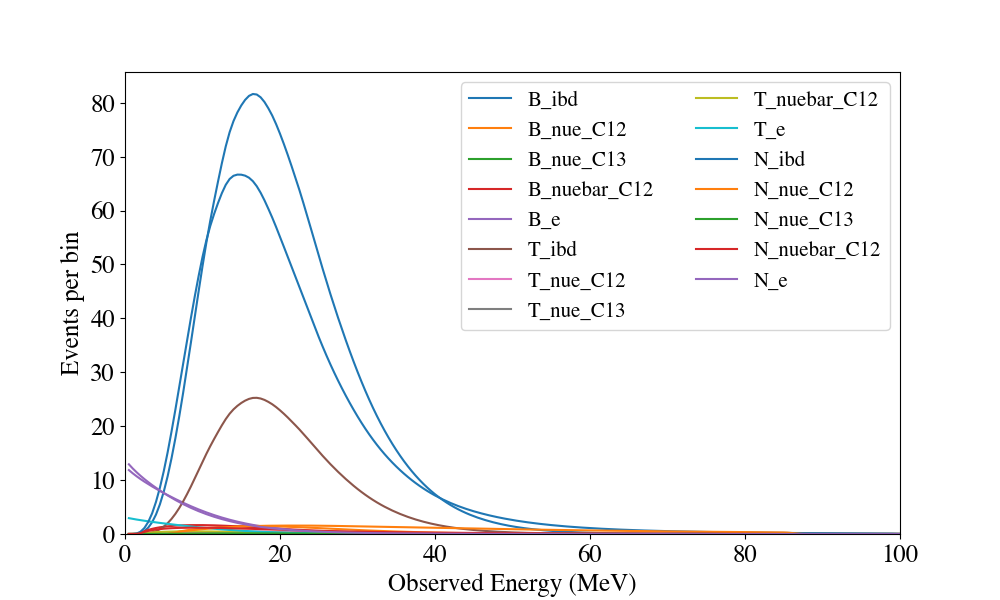}
    \caption{Event rates of JUNO where the left plot shows us the event rates without smearing being applied whereas the right plot shows event rates with smearing}
    \label{fig:eventrateJUNO}
\end{figure}

\end{itemize}

\section{Results and discussion}

\begin{figure}[h!]
    \hspace*{-1.2em}
    \includegraphics[width=0.35\textwidth]{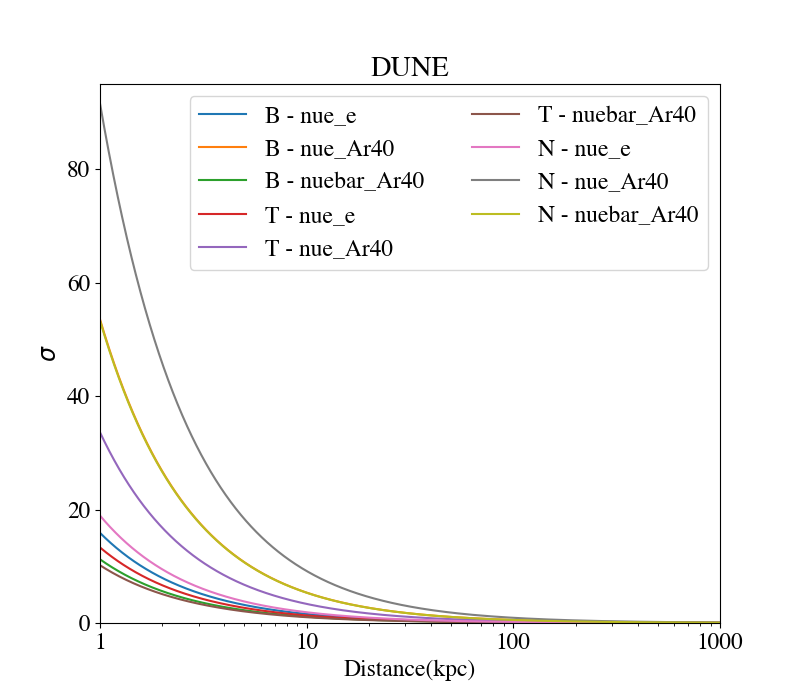}
    \hspace*{-3.5em}
    \qquad
    \includegraphics[width=0.35\textwidth]{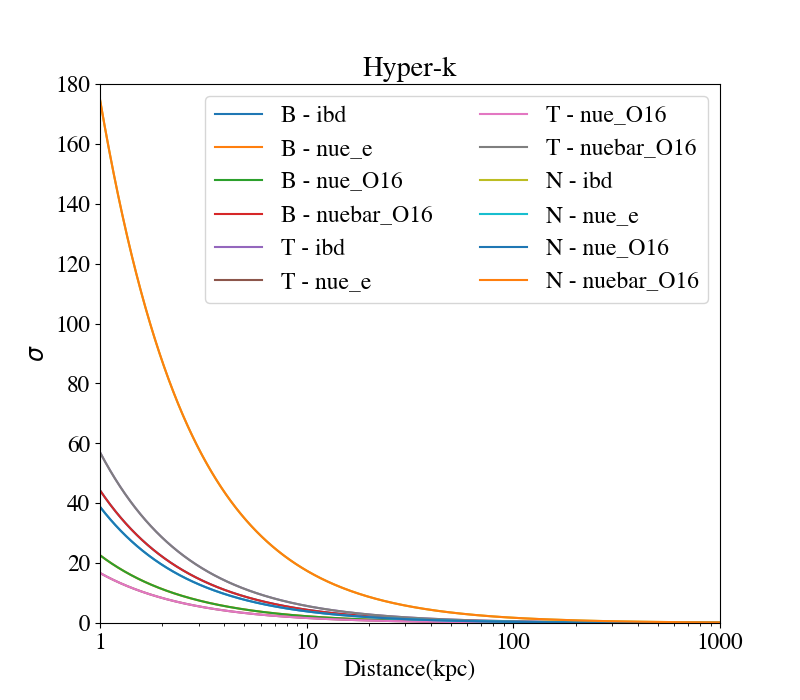} 
     \hspace*{-3.5em}
     \qquad
    \includegraphics[width=0.35\textwidth]{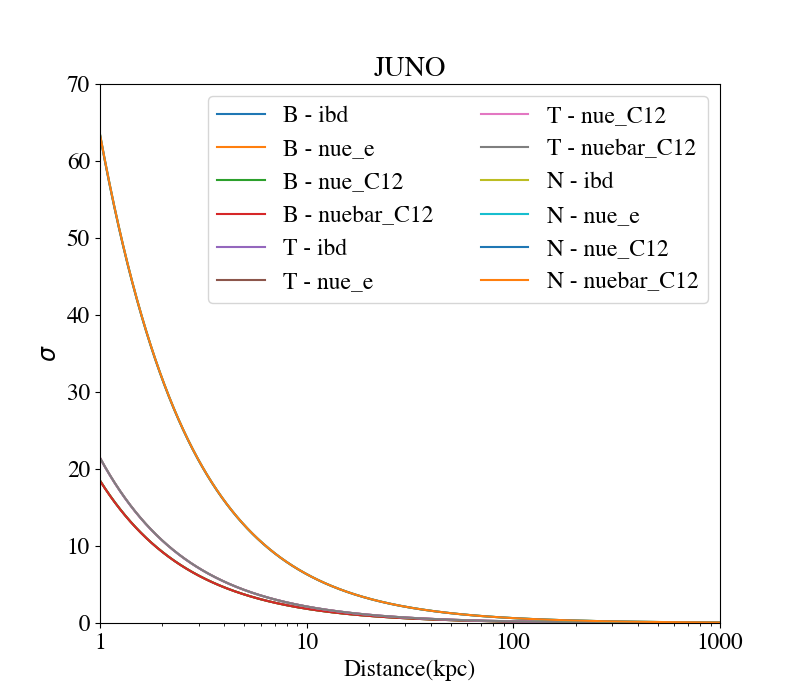}
    \caption{Sensitivity of SN neutrinos towards mass ordering in detectors without smearing being applied}
    \label{fig:sigma_unsmear}
\end{figure}

\begin{figure}[h!]
    \hspace*{-1.2em}
    \includegraphics[width=0.35\textwidth]{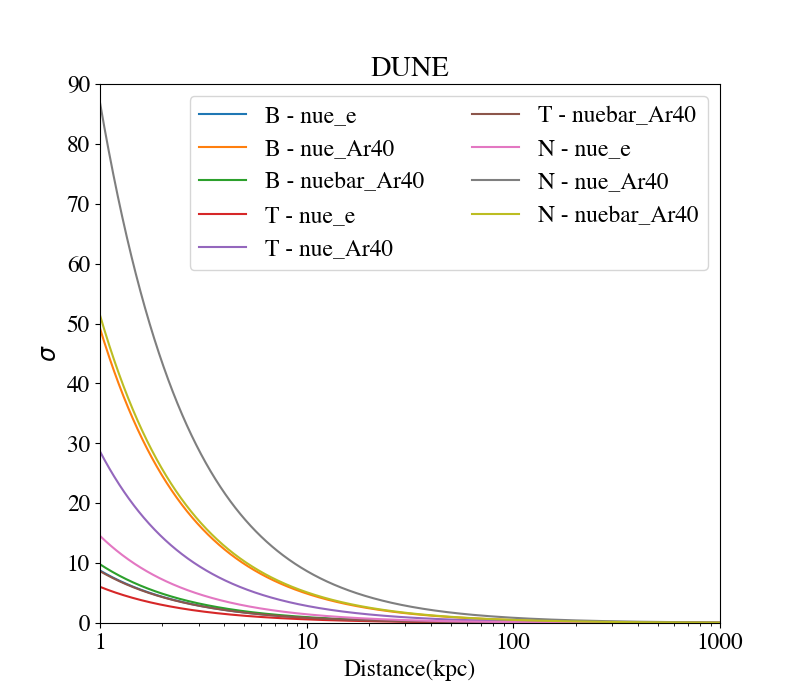}
     \hspace*{-3.5em}
    \qquad
    \includegraphics[width=0.35\textwidth]{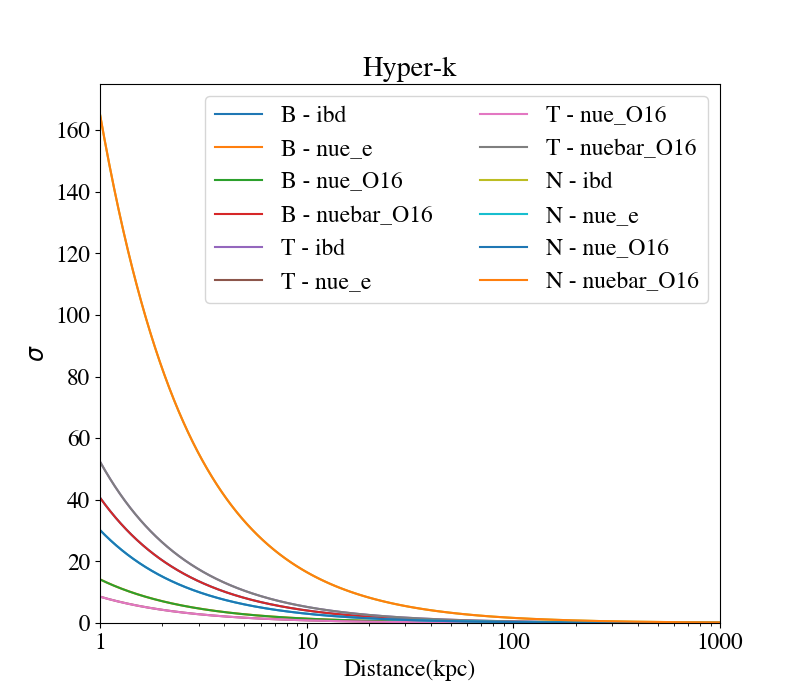}
     \hspace*{-3.5em}
     \qquad
    \includegraphics[width=0.35\textwidth]{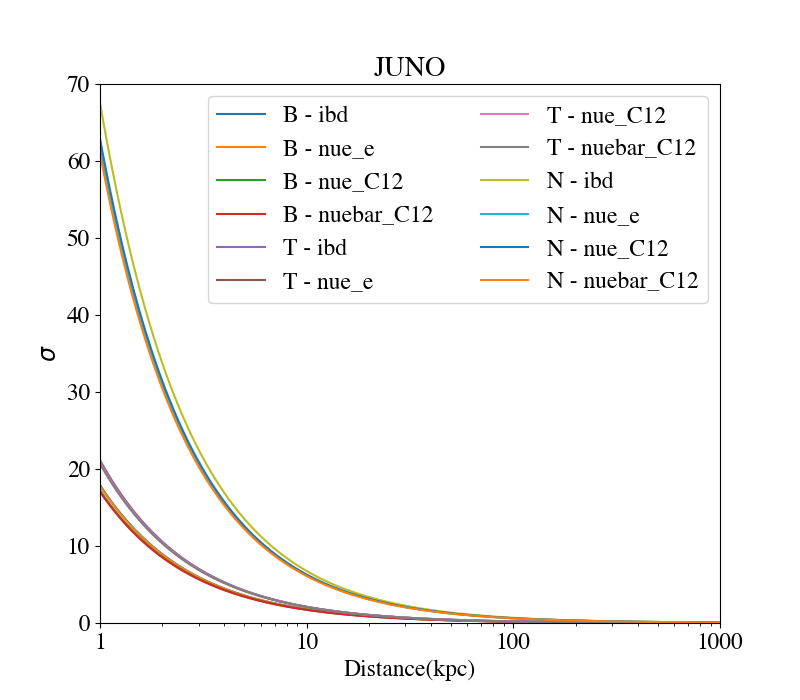}
    \caption{Sensitivity of SN neutrinos towards mass ordering in detectors with smearing being applied}
    \label{fig:sigma_smear}
\end{figure}

In this paper, to study mass ordering sensitivity towards supernova neutrinos from core collapse supernova burst, we first calculated event rates for the channels explained in the previous section. Then we calculated statistical $\chi^2$ using Poisson log likelihood\cite{gg} instead of gaussian which is usually used, as we know the number of events here are small compared to high energy neutrinos from various sources and it is given by,

\begin{equation} \label{chi2}
    {\chi^2}_{stat} = 2  \sum_{i=1}^n \biggr[ N_i^{IMO} - N_i^{NMO} - N_i^{NMO}log(\frac{N_i^{IMO}}{N_i^{NMO}}) \biggr]
\end{equation}
 
where $N_i^{NMO}$ is event rate expected for normal mass ordering which is assumed to be true as NMO is globally preferred to be true and $N_i^{IMO}$ is event rate expected for inverted mass ordering which is being tested. 
Then $\sqrt{\chi^2}$ which is $\sigma$, is plotted against supernova distance from earth in units of kpc for two cases, one for event rates where detector smearing is applied and the other without detector smearing. The channels for which $\chi^2$ is calculated is given in Table \ref{table:channels}.
For the above mentioned channels $\sigma$ is shown in fig.\ref{fig:sigma_unsmear} and \ref{fig:sigma_smear}, where fig.\ref{fig:sigma_unsmear} depicts sensitivity of DUNE, Hyper-k and JUNO without including detector smearing whereas fig.\ref{fig:sigma_smear} depicts the same when detector smearing is applied.

\begin{table}[]
    \centering
    \begin{tabular}{|c|c|c|}
    \hline
   \bf DUNE & \bf Hyper-k & \bf JUNO \\
      \hline    
    nue-e & nue-e & nue-e \\
     \hline
     nue-Ar40 & nue-O16 & nue-C12 \\
     \hline
     nuebar-Ar40 & nuebar-O16 & nuebar-C12 \\
     \hline
       & IBD & IBD \\
     \hline
    \end{tabular}
    \caption{Interaction channels studied for three detectors mentioned in the table.}
    \label{table:channels}
\end{table}

From comparing sensitivity plots for effect of detector smearing, we can clearly see that sensitivity decreases when smearing is applied which seem to be logical as there will be some limitations in detectors. At 10kpc which is considered as average Supernova distance, there is a decrease of almost 0.5$\sigma$ for nue-Ar40 in DUNE for all the flux models but in case of Hyper-k the shift in sensitivity is variable in 3 models varying from 0.3$\sigma$ to 0.7$\sigma$. For the case of JUNO, due to assumption of 100\% efficiency at present, smearing effect is very small, about 0.05$\sigma$.

We can also see here that $\sigma$ decreases with distance as number of events follow the inverse square law w.r.t distance. This diminishes the ability of detectors to distinguish between normal and inverse mass ordering at larger distances. The number of events which is expected to be roughly 3000 for DUNE, 30000 for Hyper-k and 4000 for JUNO at 10kpc which will decrease to a factor of 25 at 50 kpc. But still some existing and future detectors are massive enough to detect supernova neutrinos but their sensitivity to distinguish between both mass ordering will be greatly reduced and be close to zero above 50kpc. 

Overall, For DUNE, we can expect sensitivity to be 3$\sigma$ to 9$\sigma$ for nue-Ar40 interaction channel which is known to be its dominant channel at a distance of 10kpc according to three flux models studied here and can have sensitivity of 1$\sigma$ even at 50kpc. For other two channels, nue-e and nuebar-Ar40, sensitivity is very small and can be useful at smaller distances, like at 5kpc sensitivity of DUNE is 2$\sigma$ for Bollig and Tamborra model whereas Nakazato model seems to have sensitivity of 10$\sigma$ as shown in fig.\ref{fig:sigma_unsmear} and \ref{fig:sigma_smear}. 

However, Hyper-k and JUNO detector being sensitive to $\bar{\nu_e}$'s where IBD is seen to be a dominant channel(fig.\ref{fig:EventRateHyperk} and \ref{fig:eventrateJUNO}) have sensitivity of about 4$\sigma$ to 16$\sigma$ and 1.7$\sigma$ to 6.7$\sigma$ for different models respectively at a distance of 10kpc as shown in fig.\ref{fig:sigma_smear}. Electron flavor neutrinos seems to have very less sensitivity in these two detectors. Although event rate for $\bar{\nu_e}$ with oxygen and carbon in Hyper-k and JUNO respectively, is small compared to IBD but shows sensitivity similar to that of IBD.

\section{Summary and Future prospects}

Supernova neutrinos are produced during the collapse of massive stars, carrying away a significant portion of the gravitational energy released. This study investigates the sensitivity of future neutrino detectors; Hyper-K, DUNE and JUNO to supernova neutrinos for their ability to distinguish between normal and inverted mass hierarchy. 

The study employs three different flux models (Bollig, Tamborra, and Nakazato) to analyze the expected neutrino fluxes from core-collapse supernovae and assess how effectively these detectors can capture and analyze these particles, drawing connections to key astrophysical and high-energy physics questions.
We looked into detectors and interaction of supernova neutrinos in them, for which we calculated event rates using SNOwGLoBES as event rate calculator for each channel, as shown in fig.\ref{fig:eventRateDune}, \ref{fig:EventRateHyperk} and \ref{fig:eventrateJUNO}. To study the effect of mass ordering in detector we considered Adiabatic MSW effect where oscillation frequencies are slow compared to changing density of matter through which neutrinos pass. Adiabatic MSW for two cases i.e., normal and inverted mass ordering are applied on the supernova flux. The total number of events calculated for these flux models in said detectors are shown in Table \ref{Table:event rate}. Next,  using poisson log likelihood formula in eq.\ref{chi2} we calculated $\sigma$ in every channel with supernova distance as shown in fig.\ref{fig:sigma_unsmear} and \ref{fig:sigma_smear}. We saw that a sensitivity of at least 3$\sigma$ and 4$\sigma$ can be expected in DUNE and Hyper-k considering statistical uncertainties and 5\% error in systematical uncertainties. The study also indicate that detector sensitivity diminishes with increased distance, especially under smearing effects. 

With advancement in detector technology we can now observe a supernova even hours or days before actually observing visually. This will not only allows astronomers to get their telescopes ready in the direction of supernova but also helps us to know more about stars and core collapse supernova mechanism. 
As we saw earlier that sensitivity of detectors above 50kpc will be near to zero but if we have detectors larger(like Megatons) than those mentioned here, may be our understanding will be more deepened as we will have larger statistics at our hand. Here we carried out statistical study but on including systematic uncertainties we can further explore mass ordering sensitivity for various different flux models available to make our result more accurate. 

Also another unknown among neutrino oscillation parameters which is yet to be measured precisely is CP violating phase $\delta_{cp}$, can also be studied to some extent using different assumptions in flux models because using existing flux models, sensitivity towards $\delta_{cp}$ is zero. But on considering different approach where muon and tau flavor neutrinos have different flux distribution, then some signatures can be seen even in low energy supernova neutrinos.

\section*{Acknowledgement}
This project is funded by CSIR(Council of Scientific \& Industrial Research) and DST under grant no. SR/MF/PS-01/2016-PU/G.


\newpage
\begin{thebibliography}{99}

\bibitem{a}
R. Bionta et al.,
\emph{Observation of a Neutrino Burst in Coincidence with Supernova SN 1987a in the Large Magellanic Cloud},
\emph{Phys.Rev.Lett.} {\bf 58} (1987).


\bibitem{b}
KAMIOKANDE-II Collaboration and K. Hirata et al.,
\emph{Observation of a Neutrino Burst from the Supernova SN 1987a},
\emph{Phys.Rev.Lett.} {\bf 58} (1987) pg. 1490–1493.

\bibitem{c}
B.Abi, R.Acciarri, M.Acero, G.Adamov, D.Adams,  M.Adinolfi, Z.Ahmad, J.Ahmed and T.Alion et al.,
\emph{Volume I.Introduction to DUNE" $\bf15$}(2020).


\bibitem{d}
Francesca Di Lodovico,
\emph{The Hyper-Kamiokande Experiment}
\emph{Journal of Physics: Conference Series}{\bf 888}(2017).

\bibitem{e}
JUNO Collaboration and F. An et al.,
\emph{Neutrino Physics with JUNO}
\emph{J. Phys. G.}{\bf 43} (2016)

\bibitem{f}
Scott M.Adams, C. S.Kochanek, John F.Beacom, and Mark R.Vagins and K. Z.Stanek,
\emph{Observing The Next Galactic Supernova}
\emph{The Astrophysical Journal}{\bf 778}(2013) pg. 164.
doi: 10.1088/0004-637x/778/2/164


\bibitem{g}
Alessandro Mirizzi, Irene Tamborra, Hans-Thomas Janka, Ninetta Saviano, and Kate Scholberg, Robert Bollig, Lorenz Hudepohl and  Sovan Chakraborty,
\emph{Supernova Neutrinos: Production, Oscillations and Detection}
\emph{astro-ph.HE}{\bf 39}(2016) pg. 1-112
doi: 10.1393/ncr/i2016-10120-8

\bibitem{h}
F. Vissani,
\emph{Comparative analysis of SN1987A antineutrino fluence}
\emph{J. Phys. G: Nucl. Part. Phys.}{\bf42}(2015)


\bibitem{i}
Kate Scholberg, 
\emph{Supernova signatures of neutrino mass ordering}
\emph{Journal of Physics G: Nuclear and Particle Physics} {\bf 45}(2017)
doi:10.1088/1361-6471/aa97be

\bibitem{j}
Gianluigi Fogli, Eligio Lisi, Antonio Marrone and Irene Tamborra,
\emph{Supernova neutrino three-flavor evolution with dominant collective effects},
\emph{Journal of Cosmology and Astroparticle Physics} (2009) pg.030
doi:10.1088/1475-7516/2009/04/030.

\bibitem{k}
 C.Lunardini and A.Yu.Smirnov, 
\emph{Supernova neutrinos: Earth matter effects and neutrino mass spectrum},
\emph{Nuclear Physics B}{\bf 616} (2001) pg. 307-348.



\bibitem{l}
 K.Takahashi and K.Sato, 
\emph{Effects of Neutrino Oscillation on Supernova Neutrino -Inverted Mass Hierarchy}
\emph{Progress of Theoretical Physics}{\bf 109}(2003)pg. 919-931.



\bibitem{m}
Amanda L. Baxter, Segev Benzvi, Joahan Castaneda Jaimes, Alexis Coleiro and Marta Colomer Molla et al.
\emph{SNEWPY: A Data Pipeline from Supernova Simulations to Neutrino Signals}
\emph{astro-ph.IM}(2021)

\bibitem{n}
Irene Tamborra, Georg Raffelt, Florian Hanke, Hans-Thomas Janka and Bernhard Müller, 
\emph{Neutrino emission characteristics and detection opportunities based on three-dimensional supernova simulations}
\emph{American Physical Society (APS)}{\bf 90}(2014)


\bibitem{o}
Ken’ichiro Nakazato, Kohsuke Sumiyoshi, Hideyuki Suzuki, Tomonori Totani, Hideyuki Umeda and Shoichi Yamada, 
\emph{Supernova neutrino light curves ad spectra for various progenitor stars: from core collapse to proto-neutron star cooling}
\emph{The Astrophysical Journal Supplement Series}{\bf 205}(2013)
DOI:10.1088/0067-0049/205/1/2

\bibitem{p}
Amol S.Dighe and Alexei Yu Smirnov, 
\emph{Identifying the neutrino mass spectrum from a supernova neutrino burst}
\emph{Physical Review D}{\bf 62}(2000)


\bibitem{q}
M.Tanabashi, K.Hagiwara, K.Hikasa, K.Nakamura et al.,
\emph{Review of Particle Physics},
\emph{Particle Data Group}
\emph{Phys. Rev. D}{\bf 98}(2018)



\bibitem{r}
Florian Hanke, Bernhard Müller, Annop Wongwathanarat, Andreas Marek and Hans-Thomas Janka, 
\emph{Sasi activity in three-dimensional neutrino hydrodynamics simulations of supernova cores}
\emph{The Astrophysical Journal}{\bf 770}(2013)pg. 66.


\bibitem{s}
S. E. Woosley, A.Heger and T.A.Weaver,
\emph{The evolution and explosion of massive stars},
\emph{Rev. Mod. Phys.}{\bf 74}(2002)pg.1015-1071.



\bibitem{w}
Joshua Albert, Alex Beck, Farzan Beroz, Rachel Carr, Huaiyu Duan, Alex Friedland, Nicolas Kaiser, Jim Kneller, Elise McCarthy, Alexander Moss, Diane Reitzner, Kate Scholberg, Sebastian Torres-Lara, David Webber and Roger Wendell,
\emph{https://github.com/SNOwGLoBES/snowglobes}

\bibitem{x}
B. Abi, R. Acciarri, M. A. Acero et al., DUNE Collaboration
\emph{Supernova Neutrino Burst Detection with the Deep Underground Neutrino Experiment}
\emph{Eur. Phys. J. C. }{\bf 81}(2021)

\bibitem{y}
E. Kolbe, K. Langanke, G. Martínez-Pinedo and P. Vogel,
\emph{Neutrino–nucleus reactions and nuclear structure}
\emph{Journal of Physics G: Nuclear and Particle Physics}{\bf 29}(2003)pg. 2569–2596.

\bibitem{z}
I. Gil-Botella and A. Rubbia,
\emph{Decoupling supernova and neutrino oscillation physics with LAr TPC detectors}
\emph{Journal of Cosmology and Astroparticle Physics}{\bf 2004}(2004).


\bibitem{aa}
William J Marciano and Zohreh Parsa, 
\emph{Neutrino–electron scattering theory}
\emph{Journal of Physics G: Nuclear and Particle Physics}{\bf 29}(2003)pg. 2629–2645.

\bibitem{bb}
Resso et al.,
\emph{What can we learn on supernova neutrino spectra with water Cherenkov detectors?}
\emph{Journal of Cosmology and Astroparticle Physics }{\bf 2018}(2018)




\bibitem{cc}
Alessandro Strumia and Francesco Vissani, 
\emph{Precise quasielastic neutrino/nucleon cross-section}
\emph{Physics Letters B}{\bf 564}(2003)pg. 42–54.
   

\bibitem{dd}
E.Kolbe, K.Langanke, and P.Vogel,
\emph{Estimates of weak and electromagnetic nuclear decay signatures for neutrino reactions in Super-Kamiokande}
\emph{Phys. Rev. D}{\bf 66}(2002).


\bibitem{ee}
E.Kolbe, K.Langanke, and P.Vogel,
\emph{Weak reactions on 12C within the continuum random phase approximation with partial occupancies}
  \emph{Nuclear Physics A}{\bf 652}(1999)pg. 91–100.

\bibitem{ff}
 B. Armbruster and others,
\emph{Measurement of the weak neutral current excitation C-12(nu(mu) nu'(mu))C*-12(1+,1,15.1-MeV) at E(nu(mu)) = 29.8-MeV}",
\emph{Phys. Lett. B}{\bf 423}(1998)pg. 15-20.


\bibitem{gg}
Deepak Raikwal, Shandhya Choubey and Monojit Ghosh, 
\emph{Determining neutrino mass ordering with ICAL, JUNO and T2HK}
\emph{Eur. Phys. J. Plus}{\bf 138}(2023)


\end{thebibliography}
\end{document}